\documentclass[prb,twocolumn,amsmath,amssymb]{revtex4}

\usepackage{graphicx}
\usepackage{dcolumn}
\usepackage{bm}
\usepackage{SIunits}

\begin{document}
%
%
%
\newcommand{\gammaf}[0]{\ensuremath{\gamma_{\mathrm{f}}}}
\newcommand{\gammas}[0]{\ensuremath{\gamma_{\mathrm{s}}}}
\newcommand{\GammaB}[0]{\ensuremath{\Gamma_{\mathrm{B}}}}
\newcommand{\GammaBr}[0]{\ensuremath{\Gamma_{\mathrm{B,rad}}}}
\newcommand{\GammaBnr}[0]{\ensuremath{\Gamma_{\mathrm{B,nrad}}}}
\newcommand{\GammaD}[0]{\ensuremath{\Gamma_{\mathrm{D}}}}
\newcommand{\GammaDr}[0]{\ensuremath{\Gamma_{\mathrm{D,rad}}}}
\newcommand{\GammaDnr}[0]{\ensuremath{\Gamma_{\mathrm{D,nrad}}}}
\newcommand{\GammaBD}[0]{\ensuremath{\Gamma_{\mathrm{BD}}}}
\newcommand{\GammaDB}[0]{\ensuremath{\Gamma_{\mathrm{DB}}}}
\newcommand{\densB}[0]{\ensuremath{\rho_{\mathrm{B}}}}
\newcommand{\densD}[0]{\ensuremath{\rho_{\mathrm{D}}}}
\newcommand{\densg}[0]{\ensuremath{\rho_{\mathrm{g}}}}
\newcommand{\ket}[1]{\ensuremath{\left|#1\right>}}
\newcommand{\ketB}[0]{\ensuremath{\left|\mathrm{B}\right>}}
\newcommand{\ketD}[0]{\ensuremath{\left|\mathrm{D}\right>}}
\newcommand{\ketg}[0]{\ensuremath{\left|\mathrm{g}\right>}}
\newcommand{\Afast}[0]{\ensuremath{A_{\mathrm{f}}}}
\newcommand{\Aslow}[0]{\ensuremath{A_{\mathrm{s}}}}
\newcommand{\SiOx}[0]{\ensuremath{\mathrm{SiO}_x}}
\newcommand{\Sidiox}[0]{\ensuremath{\mathrm{SiO_2}}}
\newcommand{\nitrogen}[0]{\ensuremath{\mathrm{N_2}}}
\newcommand{\hydrogen}[0]{\ensuremath{\mathrm{H_2}}}

\title{Anomalous temperature dependence of the the spin-flip
  thermalization time between the dark and bright exciton states in
  silicon nanocrystals}

\author{Brian Julsgaard}\email{brianj@phys.au.dk}
\author{Ying-Wei Lu}
\author{Peter Balling}
\author{Arne Nylandsted Larsen}
\affiliation{Dept.~of Physics and Astronomy, University of Aarhus, Ny Munkegade
  120, DK-8000 Aarhus C, Denmark.}

\date{\today}

\begin{abstract}
  Silicon nanocrystals are studied by time-resolved fluorescence
  spectroscopy. After laser excitation the bright and dark exciton
  ground state levels are populated at random, but subsequently the
  decay curves reveal a thermalization between these levels. The
  characteristic thermalization time is found to be approximately 100
  ns for temperatures below 100 K and surprisingly increases for
  higher temperatures. The decay curves are analyzed using a simple
  two-state model for the bright and dark exciton ground states.
\end{abstract}

\maketitle

Since the discovery of light emission from porous silicon
\cite{Canham.ApplPhysLett.57.1046(1990)}, the structural and optical
properties of nano-structured silicon has been studied extensively.
In particular, the electron-hole exchange
interaction\cite{Bayer.PhysRevB.65.195315(2002)} is responsible for
splitting the exciton ground state into bright (radiative
recombination dipole allowed) and dark (almost dipole forbidden)
states, which was demonstrated
\cite{Calcott.JPhysCondensMatter.5.L91(1993),
  Brongersma.ApplPhysLett.76.351(2000),
  Vinciguerra.JApplPhys.87.8165(2000),
  Luttjohann.EurophysLett.79.37002(2007)} by time-resolved
fluorescence measurements for various temperatures while assuming the
bright and dark state populations to be in thermal
equilibrium. However, the spin-flip mechanism behind this
thermalization has not yet been studied for silicon nanocrystals
(NCs). In this letter we demonstrate that the spin-flip process can be
seen directly in the luminescence decay curves, and we measure the
characteristic time scale for the thermalization versus temperature.
\begin{figure}[b]
  \centering
  \includegraphics{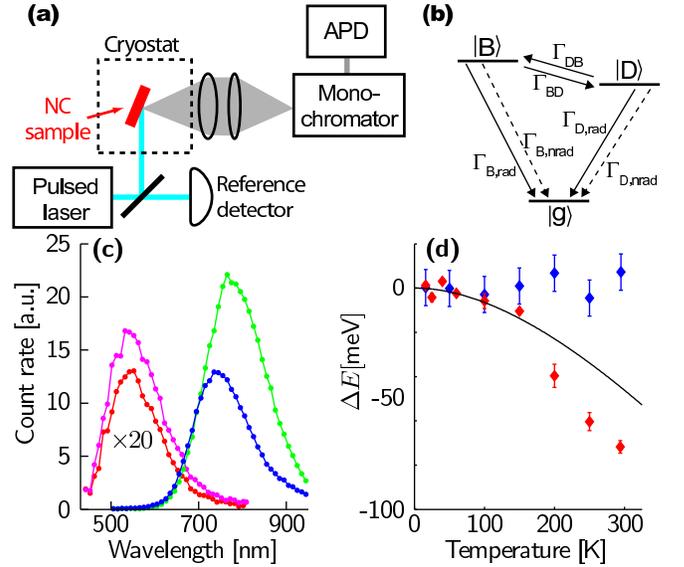}
  \caption{{\bf(a)} Experimental setup. {\bf(b)} The exciton model
    with spin-flip and decay rates defined. $\ketB$ and $\ketD$ denote
    the bright and dark exciton state, respectively, while $\ketg$ is
    the crystal ground state. ``rad'' and ``nrad'' denote radiative
    and non-radiative decay channels. {\bf(c)} Time-gated spectra. The
    early-time spectra are detected for $t < 25$ ns and shown in red
    (294 K) and magenta (16 K). The late-time spectra are detected for
    $t >100$ ns and shown in blue (16 K) and green (294 K). {\bf(d)}
    The shift in peak center position of the early-time (blue) and
    late-time (red) spectra versus temperature. The bandgap energy
    shift of bulk is shown in black.}
  \label{fig:Setup}
\end{figure}

A sample of silicon NCs was prepared by magnetron sputtering, annealed
at $1100^{\circ}\mathrm{C}$ for 1 hour in $\nitrogen$ (2 Bar), and
subsequently passivated at $500^{\circ}\mathrm{C}$ for 1 hour in 95\%
$\nitrogen$ + 5\% $\hydrogen$ (2.4 Bar). By co-sputtering Si and
$\Sidiox$ on a Si-wafer an approximately 300 nm thick layer of $\SiOx$
was achieved with $x = 1.60^{+0.05}_{-0.10}$ determined by Rutherford
backscattering spectrometry (corresponding to a silicon excess
concentration of $25^{+8}_{-4}\%$). The room-temperature luminescence
spectrum (late-time spectra in Fig.~\ref{fig:Setup}(c)) is peaked at
787 nm, being consistent with previously reported results for $\sim
2$ nm diameter NCs \cite{Iacona.JApplPhys.87.1295(2000)} (when
considering the differences\cite{Franzo.JApplPhys.104.094306(2008)}
between magnetron sputtering and plasma-enhanced chemical vapor
deposition).

\begin{figure}[t]
  \centering
  \includegraphics{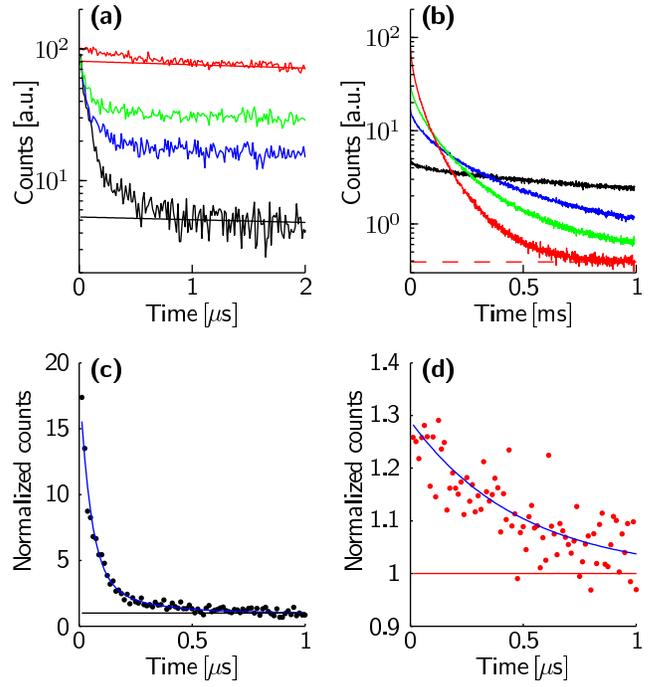}
  \caption{Decay curves for temperatures 294 K (red), 100 K (green),
    60 K (blue), and 16 K (black). The vertical axis is the same for
    all curves in {\bf(a)} and {\bf(b)}. {\bf(a)} The initial fast
    thermalization of fluorescence. {\bf(b)} The long-time decay
    curves. The red dashed line denotes the level of dark
    counts. {\bf(c)} The 16 K decay curve normalized to the black line
    of panel (a). The blue line is a bi-exponential fit. {\bf(d)} As
    panel (c) but with the 294 K data. The blue line is a
    single-exponential fit.}
  \label{fig:DecayComparison}
\end{figure}

The experimental setup is shown in Fig.~\ref{fig:Setup}(a). A
frequency-doubled Ti:sapphire femtosecond laser delivers pulses at 400
nm with a repetition rate of 1 kHz and excites the NC sample, which is
contained in a closed-cycle cryostat. The fluorescence is spectrally
filtered by a monochromator ($\Delta\lambda = 2.5$ nm) and detected by
a silicon avalanche photo-diode (APD). A reference detector enables
correction for variations in pump power.  The sample emits light in
two main bands, see Fig.~\ref{fig:Setup}(c). At early times the
luminescence is centered around 550 nm and decays on a time scale of
1.5 ns independent on emission wavelength and temperature. For this
reason, the long-wavelength tail of the early-time band can be
identified and subtracted from the late-time band, which is centered
in the range of 750-800 nm. We attribute the late-time emission to the
exciton recombination in the NCs
\cite{Calcott.JPhysCondensMatter.5.L91(1993)} while the early-time
luminescence closely resembles the emission from oxygen-related
defects \cite{Tsybeskov.PhysRevB.49.7821(1994)}. Support to this
picture is also given by the fact that the late-time emission center
energy varies with temperature in a way not very different from the
bulk silicon bandgap variation (Fig.~\ref{fig:Setup}(d)), while the
early-time band center energy is independent on temperature. In the
following we focus on the late-time band only, and the early-time
contribution has been subtracted from all the data shown in
Fig.~\ref{fig:DecayComparison}.

The ground state of the quantum confined exciton is split into two
(double-degenerate) states, a bright state, $\ketB$, and a dark state,
$\ketD$ (Fig.~\ref{fig:Setup}(b)). The bright state can recombine
radiatively via a $\Delta J = 1$ transition, while the dark state is
ideally radiatively forbidden ($\Delta J = 2$)
\cite{Bayer.PhysRevB.65.195315(2002),
  Luttjohann.EurophysLett.79.37002(2007)}. The selection rules do not
apply strictly, but we expect a small ratio, $R = \GammaDr/\GammaBr$,
between the radiative decay channels from the exciton states.

The simple model of Fig.~\ref{fig:Setup}(b) has previously been
applied successfully to other semiconductor NCs
\cite{Patton.PhysRevB.68.125316(2003),
  Labeau.PhysRevLett.90.257404(2003), Snoke.PhysRevB.70.115329(2004),
  Favero.PhysRevB.71.233304(2005), Smith.PhysRevLett.94.197402(2005),
  Johansen.arXiv.0905.4493v1} and is described by the rate equations:
\begin{equation}
\label{eq:diff_eq_exciton_states}
  \begin{split}
  \frac{\partial\densB}{\partial t} &= 
    -(\GammaB + \GammaBD)\densB + \GammaDB\densD, \\
  \frac{\partial\densD}{\partial t} &=   
    \GammaBD\densB - (\GammaD + \GammaDB)\densD,
  \end{split}
\end{equation}
where $\densB$ and $\densD$ describe the populations in the bright and
dark exciton states, respectively. The spin-flip rates between $\ketB$
and $\ketD$ are denoted by $\GammaBD$ and $\GammaDB$.  $\GammaB =
\GammaBr + \GammaBnr$ is the total decay rate of the bright state (and
similarly for $\GammaD$). Equations~(\ref{eq:diff_eq_exciton_states})
can be solved exactly, leading to a bi-exponential decay of $\densB$
and $\densD$. However, before writing the solution, we make the
following simplifying assumptions: (1) The exciton spin-flip rates are
fast compared to the exciton decay rates such that \cite{Footnote1}:
$\GammaBD$, $\GammaDB \gg \GammaB, \GammaD$. (2) The two rates
$\GammaBD$ and $\GammaDB$ must drive the populations $\densB$ and
$\densD$ toward thermal equilibrium. The relation $\GammaDB = \GammaBD
e^{-\Delta/kT}$ will assure this, where $\Delta$ is the energy
splitting between the bright and dark states, $k$ is Boltzmann's
constant, and $T$ is the temperature. (3) Immediately after laser
excitation and subsequent carrier relaxation (which takes place on a
sub-picosecond time scale \cite{Trojanek.JApplPhys.99.116108(2006)})
to the exciton ground states the population of the exciton states is
random: $\densB(0) = \densD(0) = \frac{1}{2}$. We then get:
\begin{align}
    \label{eq:rhoB}
    \densB(t) &= \frac{\frac{1}{2}[1-e^{-\Delta/kT}]e^{-\gammaf t} + 
       e^{-\Delta/kT} e^{-\gammas t}}{1+e^{-\Delta/kT}},\\
    \label{eq:rhoD}
    \densD(t) &= \frac{-\frac{1}{2}[1-e^{-\Delta/kT}]e^{-\gammaf t} + 
       e^{-\gammas t}}{1+e^{-\Delta/kT}}.
\end{align}
where the \emph{fast} decay rate characteristic of the spin
thermalization is given by: $\gammaf = \GammaBD + \GammaDB$, and the
\emph{slow} decay rate, $\gammas = \frac{\GammaB
  e^{-\Delta/kT}+\GammaD} {1+e^{-\Delta/kT}}$, characterizes the total
population decay: $\densB(t) + \densD(t) = e^{-\gammas t}$. Note that
for times, $t \gg \gammaf^{-1}$, the exciton populations have
thermalized: $\densB(t)/\densD(t) = e^{-\Delta/kT}$.

The time-dependent probability of photon emission depends on the
radiative decay rates: $p(t)= \GammaBr\densB(t) + \GammaDr\densD(t)$,
and is written:
\begin{equation}
\label{eq:p(t)}
  \begin{split}
    p(t) = &\frac{\frac{1}{2}(\GammaBr-\GammaDr)(1-e^{-\Delta/kT})}
     {1+e^{-\Delta/kT}}e^{-\gammaf t} \\
     &+ \frac{\GammaBr e^{-\Delta/kT} + \GammaDr}{1+e^{-\Delta/kT}}e^{-\gammas t} \\
     \equiv &\Afast e^{-\gammaf t} + \Aslow e^{-\gammas t}.
  \end{split}
\end{equation}
We stress that at zero time, $p(0) = \frac{\GammaBr+\GammaDr}{2}$,
independent on temperature. This reflects the initial random
population of the exciton states. At longer times, $t \gg
\gammaf^{-1}$, the second term determines the light emission.
Experimentally, the detected fluorescence from a sample containing NCs
will depend on NC density, excitation power, detection efficiency of
the entire optical setup, etc. Hence an unknown front factor must be
multiplied to Eq.~(\ref{eq:p(t)}) and we cannot determine $\GammaBr$
and $\GammaDr$ on an absolute scale but only the ratio, $R =
\GammaDr/\GammaBr$, as was also pointed out previously
\cite{Brongersma.ApplPhysLett.76.351(2000),
  Vinciguerra.JApplPhys.87.8165(2000),
  Luttjohann.EurophysLett.79.37002(2007)}. However, the ratio,
$\frac{\Afast}{\Aslow} = \frac{\frac{1}{2}(1-R)(1-e^{-\Delta/kT})} {R
  + e^{-\Delta/kT}}$, is independent on the specific experimental
setup. Information about the total decay rates, $\GammaB$ and
$\GammaD$, and the spin-flip rates, $\GammaBD$, $\GammaDB$, can be
extracted from $\gammas$ and $\gammaf$, respectively. In practice, the
decay curves will not be single-exponential due to inhomogeneous
broadening of the decay rates. However, it can easily be shown that
the predictions for $\Afast$ and $\Aslow$ are valid provided that
$\GammaBr$ and $\GammaDr$ represent the mean values of the radiative
decay rates.

Time-resolved decay curves were obtained for nine different
temperatures between 16 K and 294 K at detection wavelengths following
the center of the late-time emission spectra
(Fig.~\ref{fig:Setup}(c)). Four representative curves are shown in
Fig.~\ref{fig:DecayComparison}(a,b). In
Fig.~\ref{fig:DecayComparison}(a) it can be seen that the zero-time
fluorescence is essentially independent on temperature while it is
very different after one microsecond. This is consistent with our
assumption of initial random population in $\ketB$ and $\ketD$
followed by thermalization. In Fig.~\ref{fig:DecayComparison}(b) the
decay curves are shown for the entire laser repetition period of 1
ms. The characteristic decay time decreases with decreasing
temperature since the population freezes out
\cite{Calcott.JPhysCondensMatter.5.L91(1993),
  Brongersma.ApplPhysLett.76.351(2000),
  Vinciguerra.JApplPhys.87.8165(2000),
  Luttjohann.EurophysLett.79.37002(2007)} in the dark state,
$\ketD$. Fig.~\ref{fig:DecayComparison}(a,b) clearly demonstrate that
the spin thermalization is much faster than the exciton population
decay as was assumed in the model.  At the lowest temperatures, the
population decay time is comparable to the laser repetition period,
which must be taken into account when extracting the slow amplitude,
$\Aslow$, from these curves. The fast amplitude, $\Afast$, is
extracted by comparing the initial part of the decay curves in
Fig.~\ref{fig:DecayComparison}(a) with a local single-exponential fit
in the time range 2-5 {\micro}s (marked by straight lines for two of
the curves in Fig.~\ref{fig:DecayComparison}(a)).
\begin{figure}[t]
  \centering
  \includegraphics{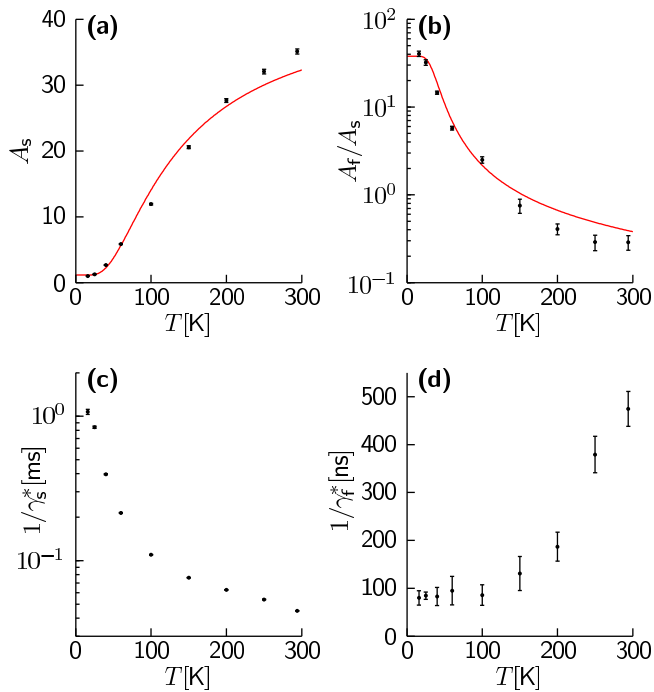}
  \caption{{\bf(a)} The measured value of $\Aslow$ versus
    temperature. The vertical axis is normalized to the 16 K data
    point. {\bf(b)} The measured ratio, $\Afast/\Aslow$, versus
    temperature. In panels (a) and (b) the red curve follows the model
    of Eq.~(\ref{eq:p(t)}) with $\Delta = 15.0\pm 1.5$ meV and $R =
    0.013 \pm 0.003$. {\bf(c)} The characteristic decay time of the
    exciton levels versus temperature. {\bf(d)} The characteristic
    time of the bright/dark-exciton state thermalization versus
    temperature.}
  \label{fig:DecayResults}
\end{figure}

The extracted values of $\Aslow$ and $\Afast/\Aslow$ are plotted in
Fig.~\ref{fig:DecayResults}(a,b) and compared to the model of
Eq.~(\ref{eq:p(t)}) (red curve). We find a reasonable agreement when
$R = 0.013\pm 0.003$ and $\Delta = 15.0\pm 1.5$ meV. The data in
Fig.~\ref{fig:DecayResults}(a) is a consequence of thermal equilibrium
and essentially confirms previously reported results
\cite{Calcott.JPhysCondensMatter.5.L91(1993),
  Brongersma.ApplPhysLett.76.351(2000),
  Vinciguerra.JApplPhys.87.8165(2000),
  Luttjohann.EurophysLett.79.37002(2007)}. The observation that the
data in Fig.~\ref{fig:DecayResults}(b) follows our model is strongly
supporting the assumption that the initial decay
(Fig.~\ref{fig:DecayComparison}(a)) is caused by spin thermalization
between the bright and dark exciton states.

Since the decay curves shown in Fig.~\ref{fig:DecayComparison}(a,b)
are in general not single exponential, we make a multi-exponential
fit, $f(t) = \sum_j a_j\exp(-\gamma_j t)$, to the curves and define
the characteristic decay rate, $\gamma^* = \sum_j a_j /
\sum_j\frac{a_j}{\gamma_j}$. This decay rate corresponds to a
single-exponential decay preserving the initial amplitude and the area
under the decay curve. The data in Fig.~\ref{fig:DecayComparison}(b)
has been fitted using three terms, and the resulting characteristic
decay time, $1/\gammas^*$, is plotted in
Fig.~\ref{fig:DecayResults}(c). The data ranges from 45 {\micro}s at
294 K to 1.1 ms at 16 K. The fact that the relative change in
$\gammas^*$, which depends on $\GammaB$ and $\GammaD$, is comparable
to the relative change in $\Aslow$, which depends on $\GammaBr$ and
$\GammaDr$, indicates that the quantum efficiency of the exciton light
emission is relatively high.

The initial thermalization part of Fig.~\ref{fig:DecayComparison}(a)
is fitted using one or two exponential terms, two examples are shown
in Fig.~\ref{fig:DecayComparison}(c,d) for 16 K and 294 K,
respectively. As exemplified, the lowest temperatures require two
terms in the fit, while one term is sufficient for the highest
temperatures. The characteristic thermalization time, $1/\gammaf^*$,
is plotted in Fig.~\ref{fig:DecayResults}(d), and we see that for low
temperatures the thermalization time is approximately 100 ns. For
higher temperatures, the initial random population is much closer to
thermal equilibrium, which seems to slow down the thermalization
rate. This slowing down is also indicated by the requirement of a
bi-exponential fit in Fig.~\ref{fig:DecayComparison}(c), although we
cannot exclude inhomogeneous broadening effects on the spin-flip time.
In the literature, a commonly applied model
\cite{Favero.PhysRevB.71.233304(2005),
  Patton.PhysRevB.68.125316(2003)} for the spin-flip rate is:
$\GammaBD = \Gamma_0 (N+1)$, $\GammaDB = \Gamma_0 N$, where $\Gamma_0$
is a characteristic zero-temperature rate and $N =
(e^{\Delta/kT}-1)^{-1}$ is the number of phonons available at the
transition energy, $\Delta$. This model is certainly invalid in our
case. Although the two-state model in Fig.~\ref{fig:Setup}(b) assumes
the (unknown) splitting \cite{Efros.PhysRevB.54.4843(1996)}
between heavy and light holes to be larger than $kT$, the analysis
shows that the model captures the main characteristics of the
spin-flip dynamics.

The results can be compared to bright/dark-state spin-flip times in
other NC systems at low temperatures. For
InAs\cite{Johansen.arXiv.0905.4493v1} and
InGaAs\cite{Favero.PhysRevB.71.233304(2005),
  Smith.PhysRevLett.94.197402(2005)} NCs similar time scales of the
order of 100 ns have been reported. However, this is not unique for
all NCs. In CdSe\cite{Patton.PhysRevB.68.125316(2003)} NCs the
spin-flip time is of the order of 10 ns, whereas a much faster upper
bound of 200 ps was reported\cite{Snoke.PhysRevB.70.115329(2004)} for
InP NCs (in this case a constant low-temperature spin-flip time was
also reported).

In conclusion, we have measured the spin-flip thermalization time
between bright and dark exciton states in silicon nanocrystals and
found a constant thermalization time of 100 ns below 100 K and
counter-intuitively increasing with temperature above 100 K.  This
work was supported by The Danish Council for Independent Research
$\vline $~Natural Sciences (FNU) and Technology and Production
Sciences (FTP, SERBINA project). We are grateful to Brian Bech Nielsen
for supplying the cryo cooler and for useful discussions.

%
%
%
%
%
%
%


\begin{thebibliography}{10}
\bibitem{Footnote1}
Strictly speaking, $\GammaDB\gg\GammaB$ does not hold at low temperatures but
  the impact on the data analysis is negligible.
\end{thebibliography}
\end{document}